\documentclass[aps,prc,reprint,nofootinbib,superscriptaddress]{revtex4-1}
\usepackage{epsfig}
\usepackage{multirow}
\usepackage{amsmath}
\usepackage{amssymb}
\begin{document}


\title{Study of $^{12}$C(p,p$^{\prime}\gamma$)$^{12}$C reaction}


\author{M. Dhibar}
\affiliation{Department of Physics, Indian Institute of Technology - Roorkee, 247667, INDIA}
\author{I. Mazumdar}
\email{indra@tifr.res.in}
\affiliation{Deptartment of Nuclear and Atomic Physics, 
Tata Institute of Fundamental Research, Mumbai - 400005, INDIA}
\author{G. Anil Kumar}
\affiliation{Department of Physics, Indian Institute of Technology - Roorkee, 247667, INDIA}
\author{S. P. Weppner}
\affiliation{Eckerd College; St. Petersburg, FL. 33711, USA}
\author{A. K. Rhine Kumar}
\affiliation{Deptartment of Nuclear and Atomic Physics, 
Tata Institute of Fundamental Research, Mumbai - 400005, INDIA}
\author{S. M. Patel}
\affiliation{Deptartment of Nuclear and Atomic Physics, 
Tata Institute of Fundamental Research, Mumbai - 400005, INDIA}
\author{P. B. Chavan}
\affiliation{Deptartment of Nuclear and Atomic Physics, 
Tata Institute of Fundamental Research, Mumbai - 400005, INDIA}


\date{\today}

\begin{abstract}
In the present work, we report our in depth study of $^{12}$C(p,p$^{\prime}\gamma$)$^{12}$C reaction both experimentally and theoretically with proton beam energy ranging from 8 MeV to 22 MeV.  The angular distributions were measured at six different angles. We discuss the gamma angular distributions, total cross sections values for 4.438, 9.64, 12.7 and 15.1 MeV states. We also describe the theoretical interpretation of our measurements using optical model analysis. We also report the branching ratios from our measurements. For the first time, we have measured the the cross section and branching ratio for the 9.64 MeV state. 
\end{abstract}


\maketitle

\section {Introduction}

\label{sec:2}
\begin{figure}[ht]
\centering
\includegraphics[height=6.cm, clip,width=0.48\textwidth]{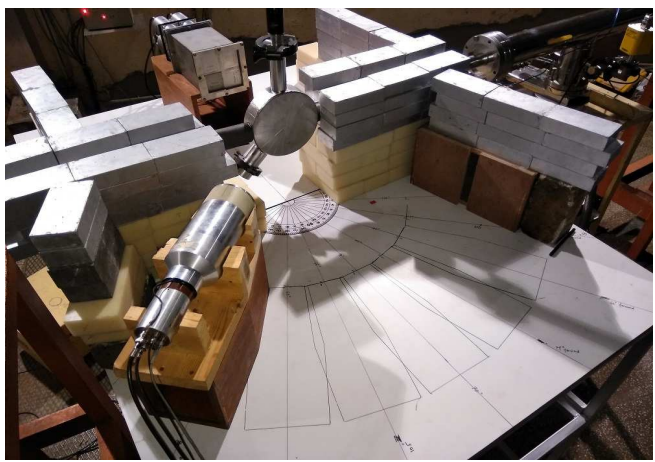}
\caption{\label{fig: 1}Experimental arrangement used during $^{12}$C(p,p$^{\prime}\gamma$)$^{12}$C
measurements.}
\end{figure}
Study of the $^{12}$C(p,p$^{\prime}\gamma$)$^{12}$C inelastic scattering reaction touches on important topics in nuclear reactions and structural studies. Understanding this reaction in terms of scattering theory involves two steps: first, the excitation, then by the de-excitation of the $^{12}$C nucleus. These steps are not equally understood.
Deexcitation process is relatively simpler to understand. It is the release of energy via a gamma ray with a known energy and angular momentum. However, understanding the excitation mechanism raises a variety of questions. Is the state under consideration excited into a collective mode or is it better understood by a single nucleon excitation, or do both representations have validity? How important is the exchange process for identical particles? The role of alpha particle clustering in $^{12}$C nucleus is also to be understood? The influence of giant resonance states in $^{12}$C and energy levels of $^{13}$N may be important in the analysis. All of these interesting questions motivate us experimentally and theoretically to study the $^{12}$C(p,p$^{\prime}\gamma$)$^{12}$C reaction with proton beam energy from 8 MeV to 22 MeV.\\ As of today, no optical model theory exists which can successfully explain excited states (namely, 4.43 MeV, 9.64 MeV, 12.71 MeV and 15.1 MeV) of $^{12}$C for this beam energy range. The lowest projectile energy that has been discussed in the literature is 40 MeV \cite{Amos2002} .  There were few attempts to formulate a microscopic proton-nucleus theory below this energy range, specifically by Jeukenne, Lejeune and Mahaux \cite{Jeukenne1976} and Brieva  Rook\cite{Brieva1977}. These theories were applied only to $^{12}$C(p, p) (only the elastic channel) at energies used in the present work (14 MeV - 40 MeV). However, several authors \cite{Petler1985,Kato1985} have struggled to fit the elastic data with their microscopic theories. Therefore, the scattering of proton in the energy range of 8 to 30 MeV poses challenge to theorists even though the optical model \cite{Weppner} formalism works well at higher energies. \\
Experimentally, there exist very few data for the cross section measurement of 12.71 MeV and 15.1 MeV states \cite{Warburton1962,Measday1973,BERGHOFER1976,Marcinkowski1999}. There exist no data on the cross sections for 9.64 MeV state. However, many studies on 4.43 MeV state provide cross section data \cite{Warburton1962,Dyer1981,Lang1987,Lesko1988,Kiener1998,Kiener2001,Kiener2017,Gul2011,Barker1963,Daehnick1964,Swint1966,Conzett1957,Guratzsch1969,Kobayashi1970}. So, there is enough justification for the measurement of scattering cross sections of $^{12}$C(p,p$^{\prime}\gamma$)$^{12}$C reaction up to about 30 MeV.\\
The  reaction  also has a large significance in the astrophysical context. The occurrence of gamma-ray lines at 4.43 MeV and 15.1 MeV from solar flares carries the details of the interaction mechanism through which charged particles (namely, protons
and ions) are accelerated and interact with  solar flares \cite{Lang1987,Ramaty1979,Kiener1998}. Since these gamma lines are  uninfluenced by solar magnetic fields, they carry the precise information about the features of the accelerated-particle population and also the identity of the accreted particles \cite{Chupp1987}. These gamma lines also provide the information about fundamental properties and conditions, such as, abundance of the elements, temperature and density of the solar flares ambience and accelerated ion-parameters. The flux ratio of $f_{15.1}$/$f_{4.43}$ from $^{12}$C(p,p$^{\prime}\gamma$)$^{12}$C  produced during solar flares can give the information about cross section as well as the relative isotropic abundance ratio of $^{12}$C and $^{16}$O \cite{Murphy1988,Murphy2009}.
\\
In the present work, we have measured the differential and total gamma cross sections and gamma branching ratios to the ground state for all the states studied.\\ The paper is categorizes in following sections. The experimental details are given is Section-2. In Section-3 we have described the mesurements and data analysis. Section-4 discusses the GEANT4 simulations and the effciency calculation of the detector. Details of theoretical formalism i.e optical model analysis are discussed in Section-5. In Section-6, we provide the results of  measurements along with a detailed theoretical analysis. Section-7 presents the summary.

\section{Experimental Details}

The $^{12}$C(p,p$^{\prime}\gamma$)$^{12}$C reaction was carried out by bombarding proton beam on mylar target of thickness 2.22 mg/cm$^{2}$. The proton beam was obtained from 14 MV Pelletron at TIFR, Mumbai\cite{Pelletron}. The experiment was performed
using proton beam ranging from  8 MeV to 22 MeV. The gamma rays produced from inelastic scattering
were detected using a 3.5$\,{''}$ \(\times\) 6$\,{''}$ large volume cylindrical LaBr$_{3}$:Ce detector\cite{Mazumdar2013}.  The gamma angular distributions were measured  at six different
angles (60$^\circ$, 75$^\circ$, 90$^\circ$, 105$^\circ$, 120$^\circ$ 135$^\circ$)  w.r.t to beam direction by keeping the detector at a distance 35 cm from the scattering chamber. The detector was shielded using lead bricks at all possible direction to reduce the background events produced during the experiment. The beam was stopped on a Faraday cup beyond the target and total charge was measured using a beam current integrator.  The energy loss of the beam in the target was found to be less than one MeV using SRIM code \cite{Srim}

\begin{figure}
\centering
\includegraphics[height=6.cm, clip,width=0.48\textwidth]{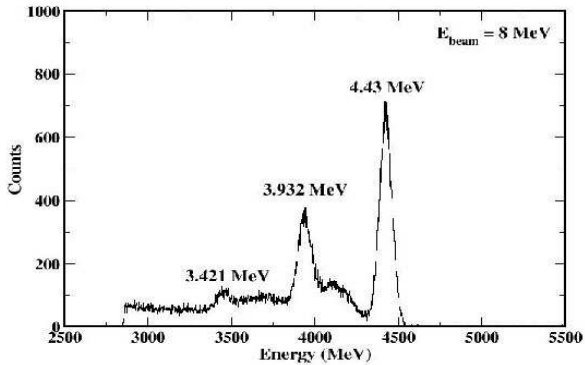}
\caption{A typical 4.438 MeV gamma spectrum from $^{12}$C(p,p$^{\prime}\gamma$)$^{12}$C
reaction at E$_p$ = 8 MeV.}
\label{4.44MeV_raw}
\end{figure}
\begin{figure}
\centering
\includegraphics[width=0.80\columnwidth]{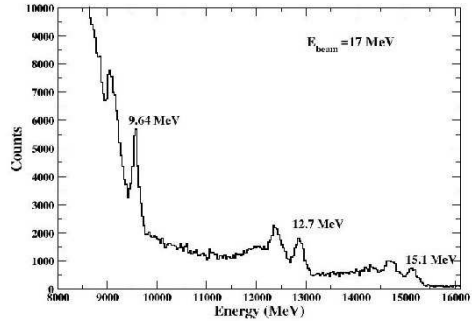}
\caption{A gamma spectrum showing 9.64, 12.7 and 15.1 MeV  from $^{12}$C(p,p$^{\prime}\gamma$)$^{12}$C reaction measured at E$_p$ = 17 MeV.}
\label{All_raw}
\end{figure}

The recorded energy and timing signals from the detector were further processed using standard Nuclear Instrumentation Module (NIM) electronics. The timing signal from anode was fed to the Time to Amplitude converter (TAC) as a start signal. Pileup  rejection  was achieved using the zero-cross technique\cite{Mazumdar1998}. The shaped and amplified energy signal was recorded in a peak sensing ADC and data were collected in events mode. Energy calibration of the detector was carried out using standard  gamma sources, namely, $^{137}$Cs (661.6 keV), $^{60}$Co (1173, 1332 keV) and $^{22}$Na (511, 1274 keV). Am-Be source was  used for 4.44 MeV gamma-rays. This ensures the energy calibration and linearity check  from few hundred keV to 4.433 MeV. The energy gain of the detector was checked  periodically during the experiment, and was found to be stable. The 4.44 MeV gamma spectrum at E$_p$ = 8 MeV is shown in Fig.2~\ref{4.44MeV_raw}.  The gamma spectrum showing three energy states of $^{12}$C nucleus, namely, 9.64, 12.7, 15.1 MeV measured  at E$_p$ = 17 MeV is presented in Fig.~\ref{All_raw}. Clearlym the use of LaBr$_{3}$:Ce detector has ensured recording spectra with quality much better than those reported using a NaI(Tl) detector \cite{Warburton1962}. The Table-\ref{T0} summarises our experimental measurements. 
\\

 \begin{figure}
\centering
\includegraphics[width=.90\columnwidth]{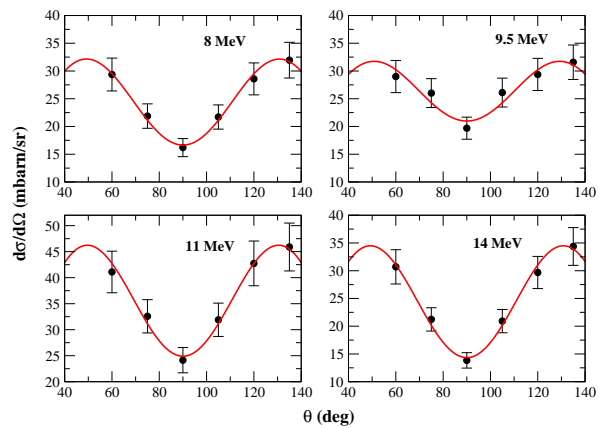}
\caption{Angular distribution of $\gamma$-rays obtained for 4.44 MeV state.}
\label{4mevdsigma}
\end{figure}
 
\section{Measurement and Analysis Method}
\label{sec:3} 
Gamma angular distributions were measured for the energy levels of $^{12}$C nucleus, namely, 4.438, 9.64, 12.71 and 15.11 MeV using inelastic scattering of protons on $^{12}$C. 
The background subtracted $\gamma$ ray yields were corrected for efficiencies of the detector for all the gamma rays to get the differential cross sections. For gamma-ray with yield $\it{Y}$,  recorded at an
angle $\theta$, with a solid angle $d\Omega$,  the differential
cross section can be written as
\begin{equation}
Y=N_{p}N_{t}(d\Omega\epsilon)\left[\frac{d\sigma}{d\Omega}\left( \theta, E_{p} \right)\right]
\label{eq1}
\end{equation}
where  $N_p$ defines the incident beam particles, $N_t$ is the number of carbon nuclei per unit area and the absolute photopeak efficiency of the detector is $\epsilon$ which is calculated using GEANT4 simulation  \cite{Geant4}. The gamma angular distribution of differential cross section was fitted using Legendre polynomial functions as given below
\begin{equation}
\frac{d\sigma}{d\Omega}=\sum ^{l=l_{max}}_{l=0;\; l=even}a_{l}P_{l}\left( cos\theta \right)
\label{eq2}
\end{equation}
where the first term a$_l$ is a free parameter obtained by
fitting data with the eq. \ref{eq2}. The second term is the Legendre polynomial function with order $l$. The $l_{max}$ should be less than  two times the gamma multipolarity. The total gamma cross section can be obtained by integrating eq. \ref{eq2} for the differential cross section to obtain 4$\pi$$a_0$, where $a_0$ is the  coefficient of the Legendre polynomial function after fitting the differential cross sections. Therefore, the total cross section can be written as,
\begin{equation}
\sigma_{total}=4{\pi}a_0
\end{equation}

\section{Simulation}
\label{sec:4}
The total detection efficiency of the detector for gamma rays  is the product of the intrinsic and geometric efficiencies. In order to make the simulations more realistic, we have included the detector geometry, the wooden support structure for the detector and all the absorbing materials between the target and the face of the detector. The low energy EM Physics package\cite{Chauvie2004} was used in the physics list class. The radioactive decay module has been used to simulate the decay of the point sources, namely, $^{137}$Cs (661.6 keV), $^{60}$Co(1173, 1332 keV), $^{22}$Na(511, 1274 keV), Am-Be(4.438 MeV) and 5.5 MeV $\gamma$-rays \cite{Hauf2013}.  The energies and branching ratios of the $\gamma$-sources used in the simulation were taken from NNDC website\cite{nndc}. The simulations were done considerng an isotropic emission of $\gamma$-rays by assuming a point source. Total sample of 10$^6$ $\gamma$-ray events were generated and distributed over the detector surface. \\

\begin{table*}
\centering
\caption{ Summary of the present measurements for 4.44, 9.64, 12.7
and 15.1 MeV states. }
\vspace{5mm}
\begin{tabular}{|c||c|c|}
 \hline
State & Beam Energy &  Angle \\
(MeV) & (MeV) &($\theta$)\\
 \hline
4.43& 8  &   60$^\circ$,75$^\circ$,90$^\circ$,105$^\circ$,120$^\circ$,135$^\circ$  \\
    &8.5& 60$^\circ$,75$^\circ$,90$^\circ$,105$^\circ$,120$^\circ$,135$^\circ$\\ 
    &9& 60$^\circ$,75$^\circ$,90$^\circ$,105$^\circ$,120$^\circ$,135$^\circ$\\     &9.5& 60$^\circ$,75$^\circ$,90$^\circ$,105$^\circ$,120$^\circ$,135$^\circ$\\
    &10& 60$^\circ$,75$^\circ$,90$^\circ$,105$^\circ$,120$^\circ$,135$^\circ$\\
    &11& 60$^\circ$,75$^\circ$,90$^\circ$,105$^\circ$,120$^\circ$,135$^\circ$\\
    &12& 60$^\circ$,75$^\circ$,90$^\circ$,105$^\circ$,120$^\circ$,135$^\circ$\\     &13& 60$^\circ$,75$^\circ$,90$^\circ$,105$^\circ$,120$^\circ$,135$^\circ$\\     &14& 60$^\circ$,75$^\circ$,90$^\circ$,105$^\circ$,120$^\circ$,135$^\circ$\\ \hline 
9.64& 14& 60$^\circ$,75$^\circ$,90$^\circ$,105$^\circ$,120$^\circ$,135$^\circ$\\
    & 15& 60$^\circ$,75$^\circ$,90$^\circ$,105$^\circ$,120$^\circ$,135$^\circ$\\
    & 16& 60$^\circ$,75$^\circ$,90$^\circ$,105$^\circ$,120$^\circ$,135$^\circ$\\
    & 17& 60$^\circ$,75$^\circ$,90$^\circ$,105$^\circ$,120$^\circ$,135$^\circ$\\
    & 18& 60$^\circ$,75$^\circ$,90$^\circ$,105$^\circ$,120$^\circ$,135$^\circ$\\
\hline
12.71    & 15& 60$^\circ$,75$^\circ$,90$^\circ$,105$^\circ$,120$^\circ$,135$^\circ$\\
    & 16& 60$^\circ$,75$^\circ$,90$^\circ$,105$^\circ$,120$^\circ$,135$^\circ$\\
    & 17& 60$^\circ$,75$^\circ$,90$^\circ$,105$^\circ$,120$^\circ$,135$^\circ$\\
    & 18& 60$^\circ$,75$^\circ$,90$^\circ$,105$^\circ$,120$^\circ$,135$^\circ$\\

\hline
15.11 & 17& 60$^\circ$,75$^\circ$,90$^\circ$,105$^\circ$,120$^\circ$,135$^\circ$\\
      & 18& 60$^\circ$,75$^\circ$,90$^\circ$,105$^\circ$,120$^\circ$,135$^\circ$\\
      & 19.5& 45$^\circ$,60$^\circ$,75$^\circ$,90$^\circ$,105$^\circ$,120$^\circ$,135$^\circ$\\
      & 20& 90$^\circ$\\
      & 20.5& 90$^\circ$\\
      & 21& 90$^\circ$\\
      & 21.5& 90$^\circ$\\
      & 22& 90$^\circ$\\

\hline
\end{tabular}

\label{T0}
\end{table*}


\section{Results and discussion}
\label{sec:6}

We present here the results of our measurements, namely, angular distributions, differential and total cross sections and branching ratios of the 4.43, 9.64, 12.7 and 15.1 MeV states. Theoretical interpretations of the data are also presented in this section.
\subsection{Excitation function of 4.438 MeV level of $^{12}$C nucleus}
\begin{figure}[ht]
\centering
\includegraphics*[width=2.5in,angle=0]{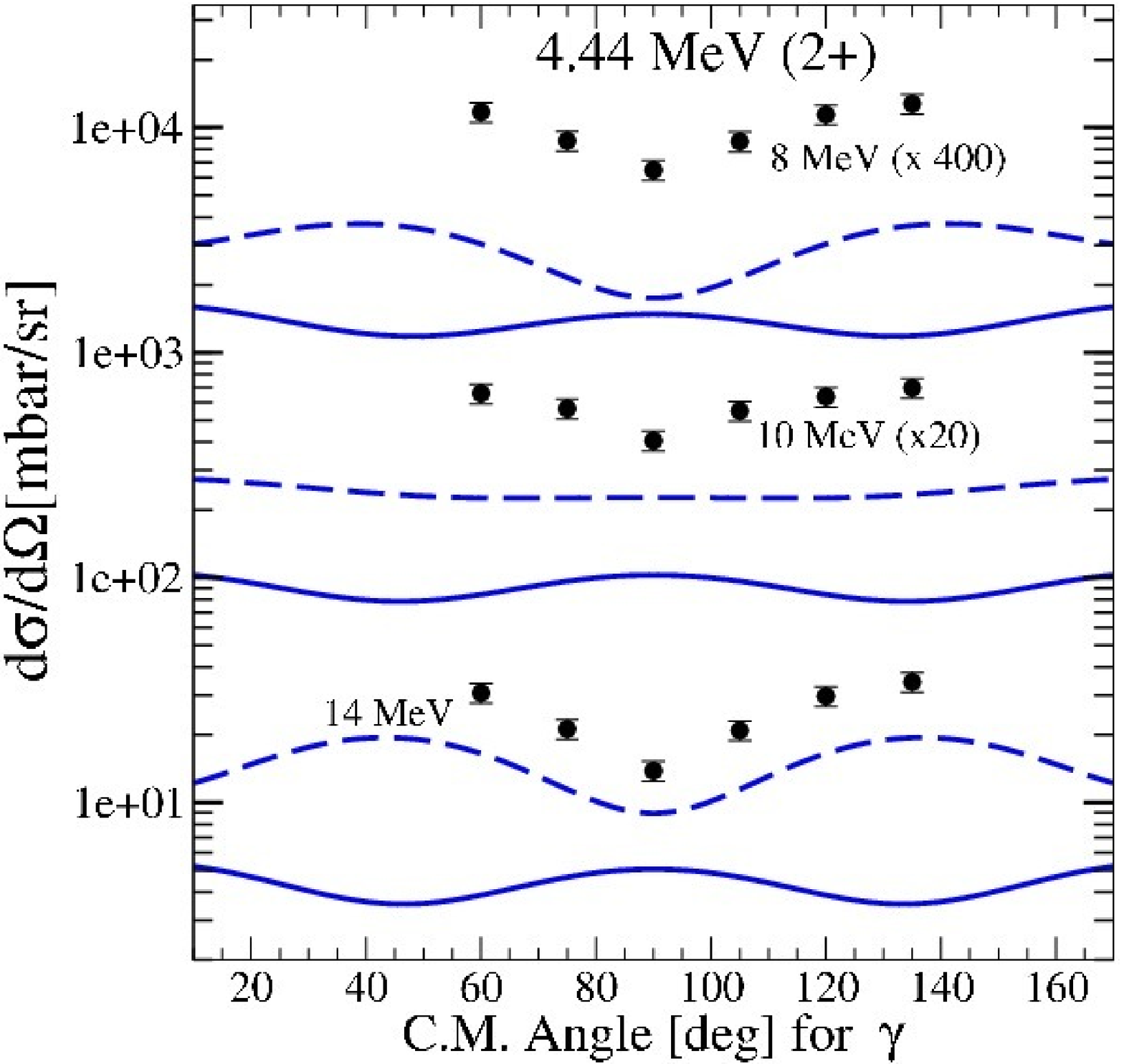}
\caption{The partial $(p,p^\prime\gamma )$differential cross-section for excitation to the 4.43 MeV (2+) energy level. The gamma data (black circle) are from the present experimental work. The blue lines are from the single particle model of this work. The solid line is the complete model without resonances added, the dashed line adds resonances listed in the Appendix \ref{appendix}.}
\label{diff_cross_section_gamma_4.43}
\end{figure}

The angular distributions and the total cross sections were measured for 4.44 MeV state at beam energies ranging from 8 to 14 MeV. A typical gamma-ray spectrum
showing the 4.44 MeV transition measured at 90$^\circ$ with respect to the
beam direction is shown in Fig.~\ref{4.44MeV_raw}. This  spectrum
corresponds to proton beam energy of 8 MeV.  The clear separation of the photopeak, first- and second-escape peaks show the high  resolving power of LaBr$_{3}$(Ce) detector. This can be contrasted with the gamma-ray spectra measured earlier with NaI(Tl) detectors with much inferior energy resolution
as discussed in \cite{Warburton1962}. Needless to say, the present measurements are aimed towards generating
many more precision data than reported earlier. Angular distributions of the gamma-rays for some of the beam energies are shown in Fig.~\ref{4mevdsigma}. The solid lines are the best fits using Legendre polynomials. 
Our method of measuring the cross sections for populating a specific state from the gamma decay of the state has potential complications. Increasing the beam energy leads to population of higher lying states above the state under consideration. Their
decay can feed to the lower state under consideration by gamma
emission, resulting in a higher observed cross section than the direct production cross section of the state under consideration. In the present case, the production cross section of the 4.44 MeV state has been measured in the rather low energy domain of 8 to 14 MeV. For these energies the population of states higher than 4.44 MeV(namely, 7.65, 9.64 and 12.7 MeV) and their subsequent feeding to 4.44 MeV is minuscule. It should be noted
that the Hoyle state at 7.65 MeV and the 9.64 MeV state predominantly decay by alpha channel. This coupled with the fact that 4.44 MeV state has  100\% gamma branching ratio to the ground state tells us that the measured cross sections of 4.44 MeV state  obtained from $(p,p^\prime\:\gamma)$ reaction
( present case) should be equal to cross sections obtained from $(p,p^\prime)$ data.\\
As discussed in the theory section there is a strong influence of
resonances in the excitation function of 4.44 MeV state. This typical structure of the excitation function is often used to verify $^{13}$N resonances in a strong coupling model~\cite{Barker1963,BARNARD1966130,BERGHOFER1976}. The
data (gamma) from Kiener $\it {et\ al.,}$ \cite{Kiener1998}  show a sharp rise in cross-section at 8.3 MeV, 9.1 MeV, 10.5 MeV, 11.0 MeV, 12.7 MeV, and 13.8 MeV. Many of these rises are at the threshold of higher energy states in the $^{12}$C nucleus ($8.3 MeV\rightarrow 7.65 MeV (0^+), 10.5 MeV \rightarrow 9.64 MeV (3^-), 12.7 MeV \rightarrow 11.8 MeV (2^-), 13.8 MeV \rightarrow 12.71 MeV (1^+)$).
The gamma differential cross sections along with  our  calculations
are shown  in Fig.~\ref{diff_cross_section_gamma_4.43}  for three  beam energies. 
The solid blue line is the single particle model without adding resonances. The single particle model with resonances (blue dashed line) appears better than the calculation without resonances in terms of  both
shape and magnitude of the data. The resonances added in the calculation are listed in Table-\ref{T5a}.  The energies, parities, and isospin ($T=0$) are taken from the compilation of Ajzenberg-Selove \cite{AJZENBERGSELOVE19901}. The theoretical calculation for 14 MeV appears to reproduce the data better
than for other two energies. The reason for this is that it is closest to the broad 15.44 MeV 2$^{+}$ resonance and therefore has a significant amount of direct excitation. In contrast the 8 MeV calculation is affected most by the 7.65 MeV 0$^+$ level (the Hoyle state) which accentuates only the exchange state. Likewise the 9.8 MeV calculation is affected by the 9.65
MeV 3$^-$ state and the 10.30 MeV 0$^+$ broad resonance which also participate in exchange. Both these calculations would benefit if there was a 2$^+$ resonance nearby to begin to replicate the signature shape of an E2 transition which all our experimental data follow. In recent times there has been a search for a 2$^+$ state as part of the Hoyle band~\cite{ITOH2004268,PhysRevC.84.054308,PhysRevC.92.021302,PhysRevLett.109.252501,Zimmerman2013} and it has been tentatively found at $(9.84 \pm 0.06)$ MeV with a width of $(1.01 \pm 0.15)$ MeV.  
\begin{figure}[ht]
\centering
\includegraphics*[width=2.5in,angle=0]{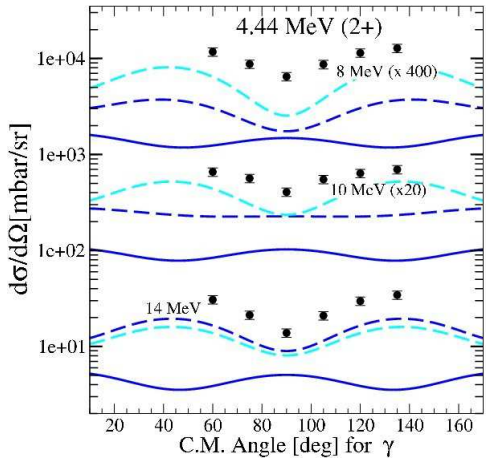}
\caption{The partial $(p,p^\prime\gamma ) $ differential cross-section of excitation to the 4.43 MeV (2+) energy level.
The gamma data (black circle) is from the present experimental work.
The blue lines are from the single particle model of this work. The solid line is the complete model without resonances added, the dashed line adds resonances listed in the Appendix \ref{appendix}. The new light blue line has the Hoyle 2$^+$ state includes as a resonance at 9.85 MeV.}
\label{diff_cross_section_gamma_4.43h}
\end{figure}
\begin{table}
\centering
\caption{Measured $\gamma$-branching ratios from $^{12}$C(p,p$^{\prime}\gamma$)$^{12}$C.}
\vspace{5mm}
\begin{tabular}{|c|c|c|c|}
\hline 
Energy&14 &18&19.5\\
(MeV)&(MeV)&(MeV)&(MeV)  \\
\hline
9.64& 0.0035$\pm$0.0031&&\\

\hline
12.7 &  &0.020$\pm$0.003&\\
\hline
15.1 &  & &0.79$\pm$0.19\\
\hline
\end{tabular}
\label{Branching}
\end{table}
 The Hoyle 2$^+$ contribution along with  present calculations of the gamma differential cross-section  is shown in Fig.~\ref{diff_cross_section_gamma_4.43h}.
The total cross section for 4.44 MeV state is shown in Fig.~\ref{cross_section_4.43h}. The experimental data are taken from  Refs.~\cite{Barker1963,PhysRev.105.1311,
BARNARD1966130,harada1,Gul2011,Daehnick1964,JPSJ.50.3198,
Conzett1957,Swint1966,Guratzsch1969,Kobayashi1970} and our data are the blue diamonds. Adding the Hoyle state  improves the total cross-section in the region of 8 to 10.5 MeV.  The gamma differential cross-section is where the calculations showed a marked difference. The shape improved dramatically, thus implying a stronger E2 transition. This shows the ability of the present calculation to translate a proton excitation to a gamma de-excitation. The single particle model has its deficits, but its microscopic power gives us strong physical insight to  the structural effect on the reaction mechanism at these low energies.
\begin{figure}[ht]
\centering
\includegraphics*[width=2.5in,angle=0]{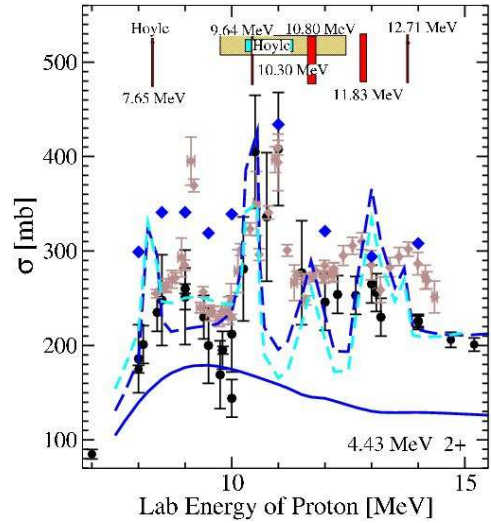}
\caption{The  total cross-section fot excitation to the 4.43 MeV (2$^+$) energy level.  The experimental data references can be found in the text. The
blue diamond data are from the present measurement. The blue lines are from the single particle model of this work. The solid line is the complete model without resonances added, the dashed line adds resonances listed in the Appendix \ref{appendix}, the new light blue dashed line includes the Hoyle 2$^+$ resonance at 9.85 MeV.}
\label{cross_section_4.43h}
\end{figure}

\subsection{Excitation function of 9.64 MeV level of $^{12}$C nucleus}
\begin{figure}
\centering
\includegraphics*[width=1.00\columnwidth]{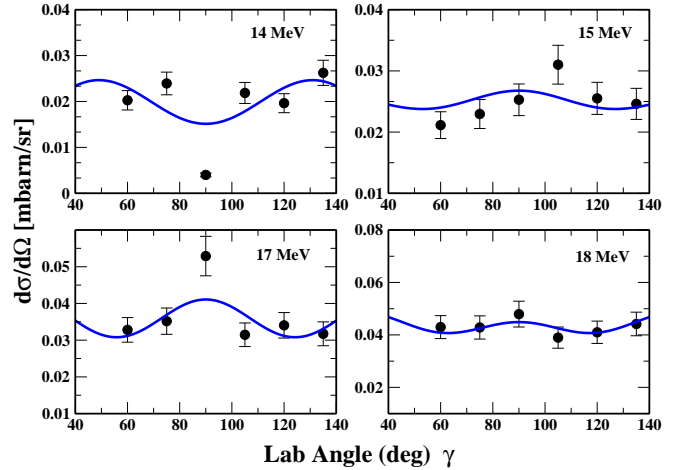}
\caption{Angular distribution of $\gamma$-rays obtained for 9.64 MeV state at E$_{p}$=14, 15,17 and 18 MeV.}
\label{9mevdsigma}
\end{figure}

\begin{figure}[ht]
\centering
\includegraphics*[width=2.5in,angle=0]{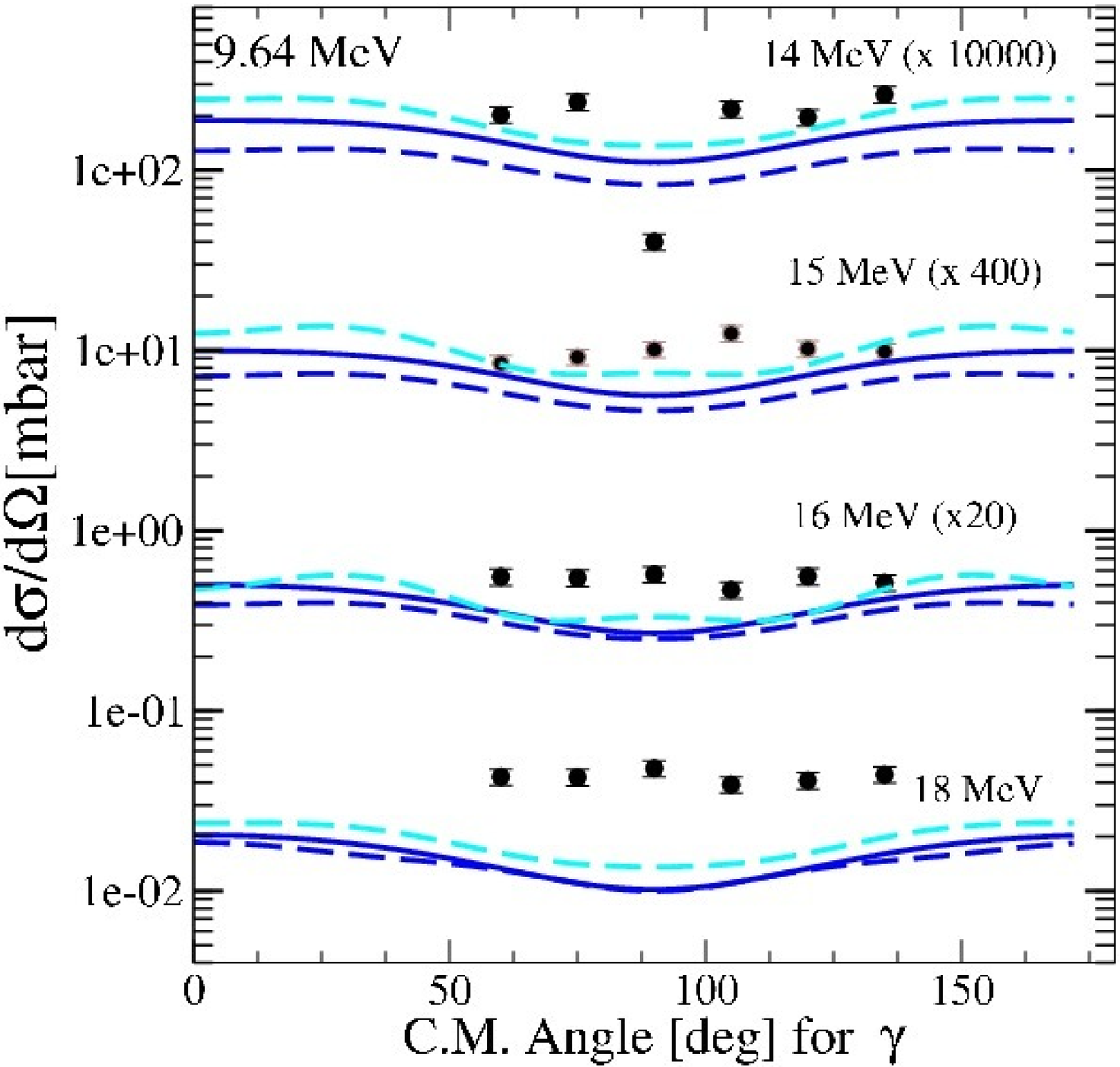}
\caption{The partial $(p,p^\prime\gamma )$ differential cross-section for excitation to the 9.64 MeV (2$^+$) energy level.
The gamma data (black circle) are from the present experimental work. The blue lines are from the single particle model of this work. The solid
line is the complete model without resonances added, the dashed line adds resonances listed in the text. The new light blue line has three additional resonances of the 3$^-$ state which are listed in the Appendix \ref{appendix}.}
\label{diff_cross_section_gamma_9.64}
\end{figure}

The 9.64 MeV state is a 3$^-$ state and also has an electric transition. The technique  used for the 4.44 MeV has been used for 9.64 MeV as well. The resonances used in the calculation are discussed in the previous section. Few additional resonances are also added to improve the results (starred symbol)
and are  shown in the same table. These resonances are taken from Refs.~\cite{PhysRevC.28.1594, AJZENBERGSELOVE19901,PhysRevC.68.014305}.
The angular distributions were measured for four beam energies ranging from 14  to 18 MeV and are shown in  Fig.~\ref{9mevdsigma}. The same measured differential cross-sections for four beam energies are shown again  in Fig.~\ref{diff_cross_section_gamma_9.64} along with
the theoretical calculation. For the lower beam energies (14 and 15 MeV) we obtained a good agreement  in terms of shape and magnitude of the data. At the highest beam energy of 18 MeV, the difference between theory and data is significant. This is because, for the 18 MeV beam energy, there may be some feeding from the higher states to the 9.64 MeV state through gamma decay. This would enhance the measured cross section of the 9.64 MeV gamma-rays.  Unlike the 4.44 MeV state
which has a 100\% gamma decay to the ground state, the 9.64 MeV state is a large alpha emitter. The value of the gamma branching ratio of the 9.64
MeV state has not previously been reported.  In this work, we have compared our measured gamma-ray cross sections at 14 MeV beam energy with the experimental  $(p,p^\prime)$ cross
section for the 9.65 MeV state reported by Harada $\it{et\ al.}$ \cite{harada1},
for the same beam energy. We have found the branching fraction to be 0.0035. This value is being reported for the first time. While inclusion of resonances
in our calculation is imported, it must be admitted that the addition of
extra E3 strengths (Table-\ref{T5b})(confusion whether to included or not??) does not lead to significant improvement
in the reproduction of the data.

\subsection{Excitation function of 12.7 and 15.1 MeV level of $^{12}$C nucleus}
\begin{figure}
\centering
\includegraphics*[width=3.5in,angle=0]{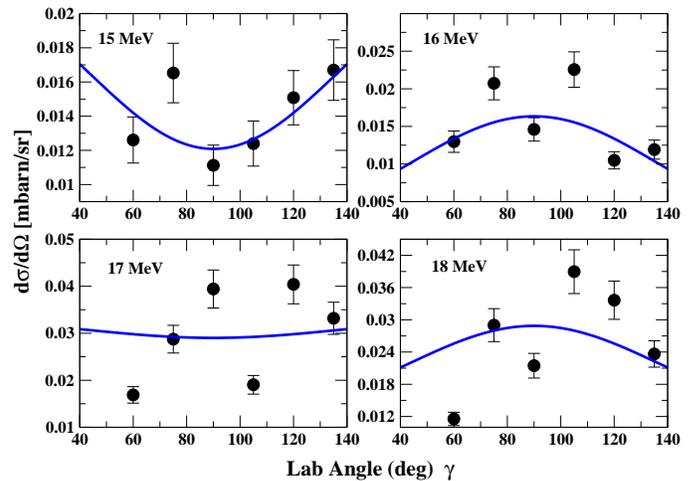}
\caption{Angular distribution of $\gamma$-rays obtained for 12.7 MeV state.}
\label{fig:12.7dsigma}
\end{figure}
\begin{figure}[ht]
        \centering
        \includegraphics*[width=2.5in,angle=0]{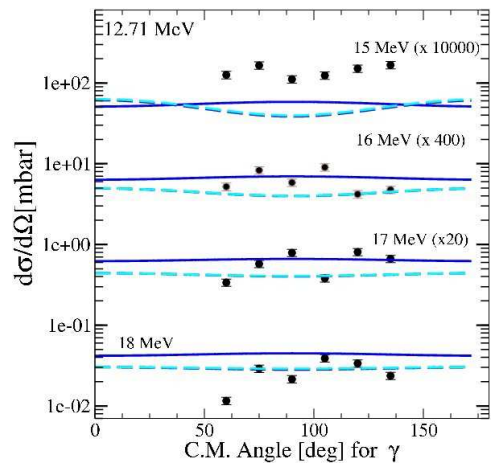}
        \caption{The partial $(p,p^\prime\gamma )$ differential cross-section for excitation to the 12.71 MeV (1$^+$) energy level. The gamma data are from this work. The blue lines are from the single particle model of this work. The solid line is the complete model without resonances added, the dashed line adds resonances listed in the Appendix \ref{appendix}. The new light blue line has
 one additional resonance of the $l=1$ state which are listed in the \ref{appendix}.}
\label{diff_cross_section_gamma_12.71}
\end{figure}
The 12.71 and 15.11 MeV states are the two lowest lying 1$^+$ states with unnatural parity of ($M1$) transitions.
   The Fig.~\ref{All_raw}. shows a typical gamma-ray spectrum measured
at 90$^{\circ}$ for a beam energy of 17 MeV. It is highly encouraging to see all the three gamma-rays namely, 9.64, 12.7 and 15.1 MeV  with their well resolved first escape peaks.  This  is much higher quality data than reoported earlier using a NaI(Tl) detector. The threshold values for the formation of 12.7 and 15.1 MeV states are 13.8 MeV, 16.8 MeV, respectively, for inelastic scattering of proton. The angular distribution of gamma-rays  were carried out for four beam energies between 15 to 18 MeV and are shown in Fig.~\ref{fig:12.7dsigma} . The solid blue lines are fits to the data with Legendre polynomials.  As mentioned in section 4.3, we extracted total cross sections from these fits.  The same experimental data (gamma differential cross sections) are compared with theoretical calculations in Fig.~\ref{diff_cross_section_gamma_12.71}. As in previous figures, the solid blue lines depict a calculation without
resonances added. The dashed blue line includes the resonances  given in Table-\ref{T5c}. The light blue dashed line also includes  two extra $l=0$ 
resonances ((1) $E=16.75 MeV, \Gamma =1.05 MeV$,Strength =$1.5E-4\;\;\;E=21.75 MeV, \Gamma =1.95 MeV$, Strength =$2.5E-4$).
Similar to the 9.64 MeV state, the 12.71 MeV state is also predominantly  an alpha emitter. The branching ratio to the ground state for gamma emission is 0.019~\cite{Marcinkowski1999,KIRSEBOM200944}.
We found the fraction to be close to $0.02$ by comparing the total $(p,p^\prime )$ cross-section and our measured $(p,p^\prime\gamma )$cross-section for this state at 18 MeV beam energy. It can be seen from Fig ~\ref{diff_cross_section_gamma_12.71}
that for  16, 17 and 18 MeV beam energies the data can be nearly reproduced by calculations with or without resonances. However, there is significant difference between theory and experiment for 15 MeV beam energy.\\

\begin{figure}
\centering
\includegraphics[width=0.8\columnwidth]{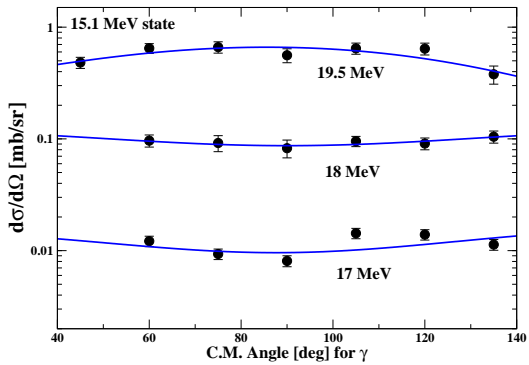}
\caption{Angular distribution of $\gamma$-rays obtained for 15.1 MeV state.}
\label{15.1disgma}
\end{figure}

\begin{figure}[ht]
\centering
\includegraphics*[width=2.5in,angle=0]{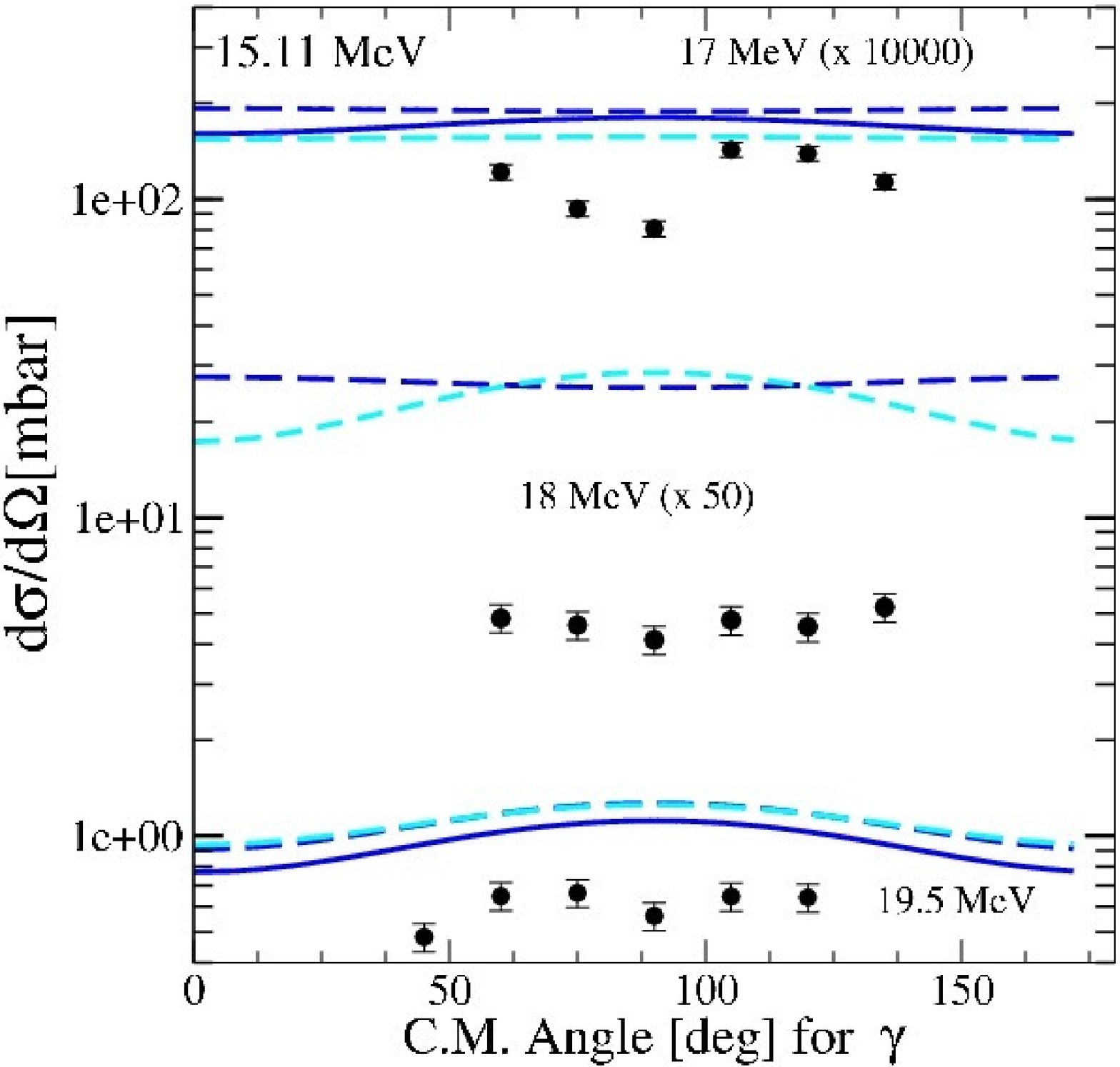}
\caption{The partial $(p,p^\prime\gamma )$ differential cross-section for excitation to the 15.1 MeV (1$^+$) energy level.
The gamma data are from this work. The blue
lines are from the single particle model of this work. The solid line is the complete model without resonances added, the dashed line adds resonances listed in the Appendix \ref{appendix}. The new light blue line has two additional resonances of the $l=0$ and $l=1$ state which are listed in the text. }
\label{diff_cross_section_gamma_15.11}
\end{figure}

For the 15.1 MeV state the 
gamma differential cross sections were measured  at three different
beam energies, namely, 17,  18 and 19.5 MeV.  The angular distributions are shown in Fig.~\ref{15.1disgma}  along with the Legendre polynomial fit (blue
line). The same gamma differential cross sections are compared with theoretical
calculations in Fig.~\ref{diff_cross_section_gamma_15.11}. We observe in Fig.~\ref{diff_cross_section_gamma_15.11}. that there is not significant difference between calculations with or without resonances. It has also to be noted that unlike the previous case of 12.71 MeV, there is significant difference in magnitudes of the experimental and theoretical cross sections. We also measured the 90$^{\circ}$ gamma differential cross sections for eight different beam energies. The data is shown in Fig.~\ref{comparision} and the inset shows the comparison of the same reported by Measday $\it{et\ al. }$ \cite{MEASDAY1963}. This shows a very good match in overall shape between our measurements and those reported in Measday $\it {et\ al,}$\cite{MEASDAY1963}.

\begin{figure}
\centering
\includegraphics[width=0.5\textwidth]{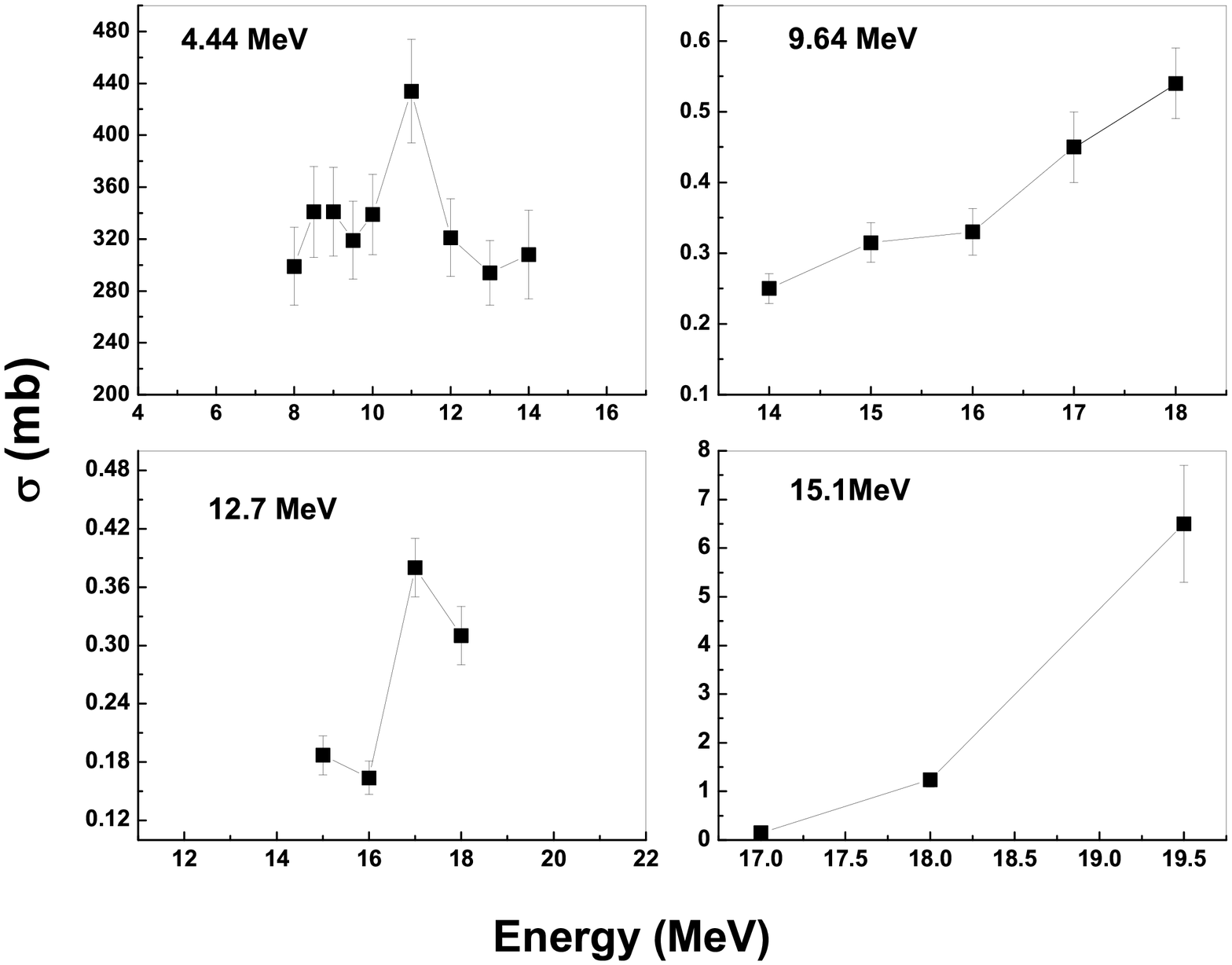}
\caption{Total measured cross section of $^{12}$C states at 4.44 MeV, 9.64 MeV, 12.7 MeV and 15.1 MeV.}
\label{totalsigma}
\end{figure}

\begin{figure}
\centering
\includegraphics*[width=0.85\columnwidth]{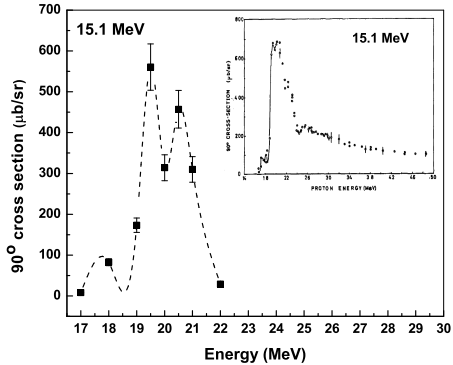}
\caption{A comparison 90$^\circ$ yield curve of present measurement of $\gamma$-rays obtained  for the 15.1 MeV state at beam energies ranging from E$_{p}$=17 to 22 MeV, and inset showing the same data of Measday $\it {et\ al,}$\cite{MEASDAY1963}.}
\label{comparision}
\end{figure}
 A good compilation of this important  data set can be found in Ref.~\cite{Marcinkowski1999}. The dashed blue lines are the theoretical calculations  carried out as a part of this work. The branching ratio to the ground state for gamma emission from the 15.1 MeV state is 0.087~\cite{Warburton1962}.
We found the fraction to be close to $0.079$ which is given in Table -\ref{Branching}. 
 \\
 The gamma breaching ratios were calculated by taking the ratio of $\sigma_{\gamma}$/$\sigma_{pp^{\prime}}$.  The values of $\sigma_{pp^{\prime}}$ cross sections  were taken
from Ref. \cite{harada1,Daehnick1964,Warburton1962}
for 9.64 , 12.71  and 15.1 MeV states. The results are
shown in Table-\ref{Branching}. We have chosen the beam energies to measure branching ratios in such a way that the desired sates (namely, 9.64 MeV, 12.7 MeV and 15.1 MeV) should not be fed significantly from higher states. Therefore, the measured branching ratios will have lesser errors. Gamma
branching ratios of 12.7 and 15.1 MeV states are in good agreement with the reported literature within the error bars. For the first time, we are reporting the gamma branching ratio of 9.64 MeV level. The total measured cross section data are shown in Fig.~\ref{totalsigma}.
\section{Summary}
\label{sec:7}
We have carried out exhaustive measurements of $(p,p^\prime\;\gamma)$ cross sections
for four excited states in $^{12}$C, (4.44, 9.64, 12.7 and 15.1 MeV) for beam energies ranging from 8 to 22 MeV. These inclided angular distribution measurements for six different angles with respect to the beam direction. The full data set is summarised in Table-\ref{T0}. Gamma-ray spectra have been measured with a large volume LaBr$_3$:Ce detector. In doing so, we have been able to record spectra with  quality better than those obtained
in the past by using NaI(Tl) detectors. We have
extracted differential and total gamma cross sections for all four states at all beam energies. Our experimental data have been compared with rigorous  theoretical optical model calculations. Overall, we have been successful in reproducing the experimental data except for the 15.1 MeV state where we found considerable mismatch between theory and experiment.
From our theoretical analysis of the data we understand it is absolutely necessary to include various resonant states in the calculations. This has been seen while reproducing the differential gamma and total gamma cross sections for all four states. We conclude that the 4.44 MeV  is a collective state which agrees with the conclusion of previous studies.
For the 9.64 MeV state, to the best of our knowledge, we have measured the cross sections and angular distributions of $(p,p^\prime\;\gamma) $  for the first time. We have also extracted the gamma branching ratio of the 9.64 MeV state to the ground state for the first time. We conclude that the 9.64 MeV state is also a collective state. The conclusions are different for the possible structures of the 12.7 and 15.1 MeV states. These two states are possibly single-particle excitations. We have been unable to reproduce the data for 15.1 MeV state as well as for the three other states. The 15.1 MeV state is especially difficult because of its unnatural $T=1$ and $M1$ transition character.
%
%
%
\section*{Acknowledgment}
One of the authors (I.M)  acknowledges the financial support from DST, Government of India,
One of the authors (M.D) acknowledges the financial support from the Ministry of Human Resource Development, Government of India, and the author (G. Anil Kumar) acknowledges the partial financial support received from DST, Government of India as part of the fast track project (No: SR/FTP/PS-032/2011).
%
%

\appendix
\section{Appendixes}
\label{appendix}

The solid blue line is the single particle model. At these energies it also is deficient. However the model has
flexibility and methods have been developed to add resonance structure to it as described above. The resonance structure
is included with the blue-dashed line. These resonances
are all the isoscalar listed in the literature~\cite{AJZENBERGSELOVE19901} but we made some choices
made when unknown. Specifically for the 18.35 MeV resonance their has been discrepancy on the
parity, we chose a $2-$ state (see Refs.~\cite{PhysRevC.28.1594,0305-4616-13-8-014}. Likewise there has been debate on what the $J$ value is
for the broad resonance at $21.6 MeV$, we chose $J\:^{\pi}=2^{+}$ following Ref.~\cite{PhysRevC.68.014305}.
Strengths of the resonance were fit to the 4.44 MeV level
data but we also used Refs.~\cite{0305-4616-13-8-014,PhysRevC.68.014305} as a guide, again with the overall goal to see
if the resonance addition help improve the calculations. Overall the result is an improvement, the best fits include
the resonance structure. 
\begin{table*}
	\centering
	\begin{tabular}{|c||c|c|c|}
		\hline
		Resonance Energy  & J$^{\pi}$ (T) & Width ($\Gamma$) & Strength \\
		MeV &  & MeV &\\
		\hline
		7.65& 0+   & 1.0E-5& 6.0E-4  \\
		9.64& 3-  & 3.4E-2 & 7.5E-4  \\
		10.80& 1-  &0.315& 1.7E-4 \\
		11.83& 2-  &0.260& 1.4E-4 \\
		12.71& 1+  &2.0E-5&1.5E-5 \\
		15.44&2+ &  1.50 & 1.25E-3 \\
		18.35&2-  &0.35  & 2.0E-5\\
		21.59&2+ & 1.20 & 4.0E-5 \\
		9.85* &   2+ &   1.01  &   2.5E-4 \\
		9.04* &  0+  &  1.45   &  7.5E-4 \\
		10.53* & 0+  &  1.42   & 1.0E-4  \\
		\hline
	\end{tabular}
	\caption{ The isoscalar resonances chosen for the light blue dashed lines for the 4.44 MeV state. The resonances were found to be the smallest values that produced a similar total cross-section for the 4.44 MeV state. The strengths are all real. The states with \* are new additions, note we removed
		the 10.3 MeV 0+ state and replaced it with two states. All electric transitions have a phase of 0, all magnetic transitions
		have a phase of $\pi$.}
	\label{T5a}
\end{table*}

\begin{table*}
	\centering
	\begin{tabular}{|c||c|c|c|}
		\hline
		Resonance Energy  & J$^{\pi}$ (T) & Width ($\Gamma$) & Strength \\
		MeV &  & MeV &\\
		\hline
		4.44&    2+&  1e-06 & 3.0E-3 \\
		7.65& 0+   & 1.0E-5& 6.0E-4  \\
		10.80& 1-  &0.315& 1.7E-4 \\
		11.83& 2-  &0.260& 1.4E-4 \\
		12.71& 1+  &2.0E-5&1.5E-5 \\
		15.44&2+ &  1.50 & 1.25E-3 \\
		18.35&2-  &0.35  & 2.0E-5\\
		21.59&2+ & 1.20 & 4.0E-5 \\
		9.85 &   2+ &   1.01  &   2.5E-4 \\
		9.04 &  0+  &  1.45   &  7.5E-4 \\
		10.53 & 0+  &  1.42   & 1.0E-4  \\
		14.25*&    3-  &  1.1 & 3.0E-4\\
		18.6* &   3-  &  0.35 & 3.8E-4\\
		21.59*&    3-&    1.2 & 1.05E-3\\
		\hline
	\end{tabular}
	\caption{ The isoscalar resonances chosen for the dark and light blue dashed lines of the 9.64 MeV state. The resonances were found to be the smallest values that produced a similar total cross-section for the 4.44 MeV state. The strengths are all real. The states with \* are new additions for the
		light blue line for the 9.64 MeV state. All electric transitions have a phase of 0, all magnetic transitions
		have a phase of $\pi$. These resonances are nearly the same as for the 4.44 MeV state.}
	\label{T5b}
\end{table*}

\begin{table*}
	\centering
	\begin{tabular}{|c||c|c|c|}
		\hline
		Resonance Energy  & J$^{\pi}$ (T) & Width ($\Gamma$) & Strength \\
		MeV &  & MeV &\\
		\hline
		4.44&    2+&  1e-06 & 3.0E-3 \\
		7.65& 0+   & 1.0E-5& 6.0E-4  \\
		9.64& 3-   & 0.034& 7.5E-4 \\
		10.80& 1-  &0.315& 1.7E-4 \\
		11.83& 2-  &0.260& 1.4E-4 \\
		15.44&2+ &  1.50 & 1.25E-3 \\
		18.35&2-  &0.35  & 2.0E-5\\
		21.59&2+ & 1.20 & 4.0E-5 \\
		9.85 &   2+ &   1.01  &   2.5E-4 \\
		9.04 &  0+  &  1.45   &  7.5E-4 \\
		10.53 & 0+  &  1.42   & 1.0E-4  \\
		14.25&    3-  &  1.1 & 3.0E-4\\
		18.6 &   3-  &  0.35 & 3.8E-4\\
		21.59&    3-&    1.2 & 1.05E-3\\
		18.16*&    1+&    0.24& 3.0E-4\\
		\hline
	\end{tabular}
	\caption{ The isoscalar resonances chosen for the dark and light blue dashed lines for the 12.71 MeV state. The resonances were found to be the smallest values that produced a similar total cross-section for the 4.44 MeV state. The strengths are all real. The states with \* are new additions for the
		light blue line for the 12.71 MeV state. All electric transitions have a phase of $\pi$, all magnetic transitions
		have a phase of 0 (since 12.71 MeV is a magnetic transition). These resonances are nearly the same as for the 4.44 MeV and 9.64 MeV states.}
	\label{T5c}
\end{table*}

\begin{table*}
	\centering
	\begin{tabular}{|c||c|c|c|}
		\hline
		Resonance Energy  & J$^{\pi}$ (T) & Width ($\Gamma$) & Strength \\
		MeV &  & MeV &\\
		\hline
		16.11  &  2+ &   5.0E-3 &  1.0E-4  \\
		16.57  &  2- &   0.3    & 3.0E-4  \\
		17.23  &  1-  &  1.15   &  7.0E-4 \\
		17.76  &  0+  &  0.08   &  6e-05  \\
		18.35  &  3-  &  0.22   &  9e-05  \\
		18.8   & 2+ &   0.1 &    8e-05    \\
		22  &  1-   & 0.8    &  1.2E-4    \\
		22.65 &   1- &   3.2  &   1.5E-4  \\
		23.53* &   1- &   0.238 &   3.0E-4\\
		23.99* &   1- &   0.57  &   5.0E-4\\
		25.4*  &  1-  & 2.0  &      1.1E-3\\
		23.75* &   0- &   3.0 &     1.5E-3\\
		\hline
	\end{tabular}
	\caption{ The isovector resonances chosen for the dark and light blue dashed lines of the isovector 15.11 MeV state.
		These resonances do not have the same confidence level as the isoscalar transitions, they were fit only to the 15.11 MeV state. The strengths are all real. The states with \* are new additions for the
		light blue line for the 15.11 MeV state. All electric transitions have a phase of $\pi$, all magnetic transitions
		have a phase of 0 (since 15.11 MeV is a magnetic transition).}
	\label{T5d}
\end{table*}

\end{document}